\documentstyle[twoside,fleqn,psfig,espcrc2]{article}

\newlength{\figwidth}
\title{Monte Carlo Renormalization Group analysis of QCD
in two dimensional coupling space%
\thanks{Poster by H. Matsufuru at LATTICE 99.}}

\author{QCD-TARO Collaboration:
Ph.~de~Forcrand%
\address{SCSC, ETH-Z\"urich, CH-8092 Z\"urich, Switzerland \\
$^{\rm b}$Dept. F\'{\i}sica Te\'orica, Universidad Aut\'onoma de Madrid,
      E-28049 Madrid, Spain \\
$^{\rm c}$Dept. of Appl. Phys., Fac. of Engineering,
           Fukui Univ., Fukui 910-8507, Japan \\
$^{\rm d}$Dept. of Physics, Tezukayama Univ.,Nara 631-8501, Japan \\
$^{\rm e}$Research Center for Nuclear Physics, Osaka Univ.,
           Ibaraki 567-0047, Japan \\ 
$^{\rm f}$Dept. of Physics, Hiroshima Univ.,
           Higashi-Hiroshima 739-8526, Japan \\
$^{\rm g}$Res. Inst. for Inform. Sc. and Education, Hiroshima Univ.,
           Higashi-Hiroshima  739-8521, Japan \\
$^{\rm h}$Institut f\"ur Theoretische Physik, Univ. Heidelberg
           D-69120 Heidelberg, Germany \\
$^{\rm i}$FEST, Schmeilweg 5, D-69118 Heidelberg, Germany \\
$^{\rm j}$Hiroshima University of Economics, Hiroshima 731-01, Japan },
M.~Garc{\'\i}a~P\'erez$^{\rm b}$,
T.~Hashimoto$^{\rm c}$,
S.~Hioki$^{\rm d}$,
H.~Matsufuru$^{\rm e}$,
O.~Miyamura$^{\rm f}$,
A.~Nakamura$^{\rm g}$,
I.-O.~Stamatescu$^{\rm hi}$,
T. Takaishi$^{\rm j}$,
and
T.~Umeda$^{\rm f}$}

\begin{document}

\begin{abstract}
We report our results of the Monte Carlo Renormalization
Group analysis in two dimensional coupling space.
The qualitative features of the RG flow are described with a
phenomenological RG equation.
The dependence on the lattice spacing for various actions provides
the conditions to determine the parameters entering the RG equation.
\end{abstract}

\maketitle

\section{Strategy and previous results}

The Monte Carlo Renormalization Group (MCRG) approach has been applied
to investigate the QCD dynamics under the renormalization transformation
and to develop improved lattice actions which enable us
to carry out simulations close to the continuum limit even
on rather course lattices.
In the multi-dimensional coupling space, an action $S$ is transformed
from one point to another point under the blocking transformation.
There is a special trajectory (renormalised trajectory: RT)
which starts from the ultra-violet fixed point and on which
actions keep long range contents corresponding to continuum physics.
Recently Hasenfratz and Niedermayer reminded us of this fact and called
the action on the RT a ``perfect action'' \cite{HN94}.

\begin{figure}[tb]
\leavevmode\psfig{file=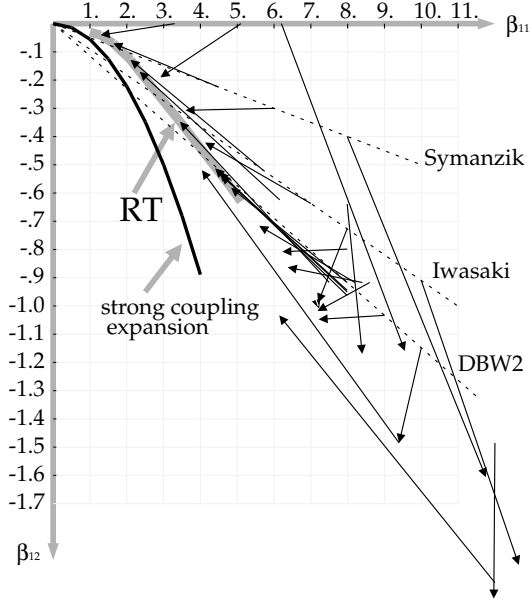,width=0.9\figwidth}
\vspace{-1.2cm} \\
\caption{Coupling flow in two-dimensional space.}
\label{fig:flow}
\end{figure}

This report summarises our previous MCRG study of QCD in two-dimensional
coupling space and proposes a model equation to describe the observed
flow features.
The action takes the form
\begin{eqnarray}
S = & &  \beta_{11}\sum_{plaq} (1-\frac{1}{3}\mbox{ReTr}U_{plaq})
 \nonumber \\
   &+& \beta_{12}\sum_{rect} (1-\frac{1}{3}\mbox{ReTr}U_{rect})
\end{eqnarray}
where ``plaq'' and ``rect'' mean $1 \times 1$ and $1 \times 2$
loops respectively.
We apply the Swendsen's blocking transformation \cite{Swe79}
and determine the couplings of resultant action $S'$ by
the Schwinger-Dyson equation method \cite{GAO87}.
In Fig. \ref{fig:flow} the coupling flow under a factor 2 blocking
transformation is expressed as arrows \cite{TARO98b}.
The observed attractive flow to a narrow stream indicates that
the RT lies near the present coupling space.
This fact was shown more manifestly by performing simulations in  the three-dimensional
coupling space which contains the 3D twist operator in addition to the plaquette
and rectangular loops \cite{TARO98a}:
Starting from the plane $(\beta_{11}, \beta_{12}, \beta_{twist}\!=\!0)$,
the departure from this plane is small and flow features are almost
the same as the ones obtained in two coupling space.

It is hence expected that actions lying on the RT of the 2D coupling space
actually reduce the finite lattice spacing effects.
To examine this, we defined the DBW2 action (doubly blocked Wilson action
in 2D coupling space) through the line defined by the points 
obtained with successive twice blocking from the Wilson action
\cite{TARO99a}.
In Fig. \ref{fig:flow}, DBW2 is represented by a dotted line
as well as Iwasaki and Symanzik actions \cite{Iwa83,Sym83}.
We analysed the rotational symmetry restoration effect
and the scaling of $T_c$ with the coupling,
and found that DBW2 actually improves these quantities.

\section{Model equation of flow}

We try to describe the observed flow in terms of a model equation
which contains a marginal and a irrelevant coupling.

Let us start in the weak coupling region.
We adopt the following equation:
\begin{eqnarray}
{d \vec{\beta} \over d {\rm ln} a} = {\bf A} \vec{\beta} - B(\beta_n)
{\vec{n} \over |\vec{n}|^2}.
\label{eq:weakRG}
\end{eqnarray}
where $\vec{\beta} ={}^t(\beta_{11}, \beta_{12})$ and $\beta_n=(\vec{n}\cdot \vec{\beta})$ .
Due to asymptotic scaling,
the constant $2 \times 2$ matrix ${\bf A}$ has eigenvalues $\lambda=0$  and
$\lambda<0$ (irrelevant) with eigenvectors $\vec{w}$ and $\vec{v}$
respectively.
$B(x)=12 b_0 + 72 b_1/x$ with
$b_0=33/(48\pi^2)$ and $b_1=(102/121) b_0^2$.
$\vec{n}$ is a constant vector orthogonal to $\vec{v}$.

In eq.(\ref{eq:weakRG}), the two-loop asymptotic scaling holds 
for $\beta_n$ as 
\begin{equation}
a\Lambda =
   ({\beta_n/6+b_1/b_0 \over b_0})^{b_1/2b_0^2} \, \exp(-\beta_n/12b_0).
\end{equation}

To obtain the model equation in the strong coupling region,
let us consider the strong coupling calculation of Wilson loops
in two coupling space.
\begin{eqnarray}
\langle W(N \times M) \rangle \hspace{-0.2cm}
  &=& \hspace{-0.2cm} (\beta_{12}/18)^{NM}
\nonumber \\
  & & \hspace{-1.3cm} + (\beta_{12}/18)^{NM-2} (\beta_{12}/18)P_1^{NM}
      + \cdots.
\end{eqnarray}
where $P_k^{NM}$ are tiling weights for filling the area by
$(NM-2k)$ [$1 \times 2$] and $k$ [$1 \times 1$] tiles.
Imposing the area law with physical string tension $\sigma$,
the $N \times M$ Wilson loop is expressed as
\begin{equation}
\langle W(N \times M) \rangle
 = \exp(-a^2 \sigma NM + b(N+M)+c).
\end{equation}
For $\beta_{12}=0$, these equations require
$\beta_{11}/18 = \exp(-a^2 \sigma)$
in the leading order of $NM$.
Then for finite $\beta_{12}$, 
$\beta_{12}/18 = C \exp(-2a^2 \sigma)$ follows to the same order.

Thus the model equation in the strong coupling region is as follows.
\begin{eqnarray}
{d \vec{\beta} \over da^2} = - 
\left(\begin{array}{cc}
 \sigma + \zeta_1/a& 0 \\ 
 0 & 2 \sigma + \zeta_2/a\\
\end{array}
\right)
\vec{\beta}\ \ ,
\label{eq:strongRG}
\end{eqnarray}
where we include the next order contributions with parameters
$\zeta_1$ and $\zeta_2$.

Combining equations (2) and (6), we come to the following RG equation
in terms of a dimensionless variable $a_s=a\sqrt{\sigma}$:
\begin{eqnarray}
{d \vec{\beta} \over da_s} = 
&-& 2  \left( \begin{array}{cc}
                a_s  + \bar{\zeta_1}& 0 \\ 
                0 & 2a_s + \bar{\zeta_2} \\
              \end{array}  \right)  \vec{\beta}
\nonumber \\
&+& {1 \over a_s } \left[ {\bf A} \vec{\beta}
                    - B(\beta_n){\vec{n} \over |\vec{n}|^2}\right]
\end{eqnarray}
where $\bar{\zeta}_i = \zeta_i/\sqrt{\sigma}$.

The two vectors $\vec{v}$ and $\vec{w}$ are parameterised 
as 
$\vec{v}=(\cos\theta',\sin\theta')$ and $\vec{w}=(\cos\theta,\sin\theta)$ .
Then the matrix ${\bf A}$ with a zero and a finite eigenvalue $\lambda$ 
is given by
\begin{eqnarray}
{\bf A} = {\lambda \over (\vec{v}\cdot\vec{w_T})}\vec{v} \otimes \vec{w_T}^t
\end{eqnarray}
where $\vec{w_T}$ is a vector orthogonal to $\vec{w}$. 
The vector $\vec{n}$ is given by $\vec{n}=(1,\cot(-\theta'))$.

\begin{figure}[tb]
\vspace*{-0.8cm}
\leavevmode\psfig{file=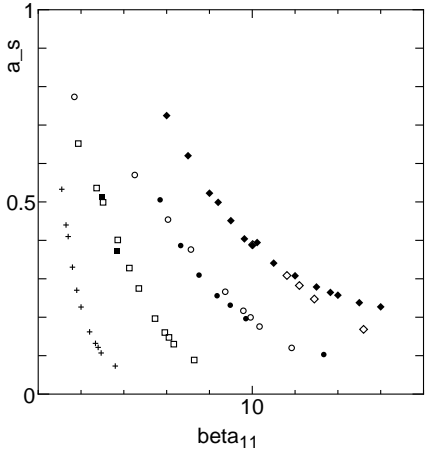,width=\figwidth}
\vspace{-2.4cm} \\
\caption{
$\beta_{11}$ vs $a_s=a\sqrt{\sigma}$.
Filled symbols: fitted data,
Open symbols: results of fit .}
\label{fig:model1}
\end{figure}

\begin{figure}[tb]
\vspace*{-0.8cm}
\leavevmode\psfig{file=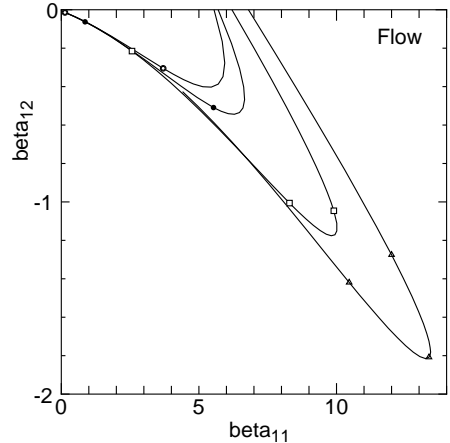,width=\figwidth}
\vspace{-2.4cm} \\
\caption{
Coupling flow described by the model equation.
Values of the parameters are found in the text.}
\label{fig:model2}
\end{figure}

We fit the results of \cite{TARO99a,BR98,CPPACS99} to this RG equation.
Preliminary results of the fit are shown in Figs. \ref{fig:model1}
and \ref{fig:model2}.
To show the qualitative features, we use
$\theta=-0.156$, $\theta'=-0.205$, $\lambda= -0.5$, 
$\bar{\zeta_1} = 0.1$ and $\bar{\zeta_2}=0.0$.
These parameter values are not the result of well-tuned fit.
Figure \ref{fig:model1} shows $\beta_{11}$ vs $a\sqrt{\sigma}$.
Filled symbols are fitted data while open symbols are results of the fit.
In Fig. \ref{fig:model2}, the flow described by the model equation
is displayed.
These results indicate that the qualitative features of the flow
are well described.

\section{Conclusions and outlook}

In this proceedings, we summarise our previous works on the
MCRG analysis in two dimensional coupling space and propose a model
equation which, although
sufficiently simple to handle, describes well the features of the observed RG flow.
A more quantitative analysis will obtain the
renormalization trajectory in two-coupling space on which
the action is expected to contain quite small finite
lattice spacing effects.

\bigskip

The calculations have been done
on AP1000 at Fujitsu Parallel Computing Research Facilities,
Fujitsu VPP/500 at KEK.
This work is supported by the Grant-in-Aide for Scientific Research 
by Monbusho, Japan (No. 11694085).

\end{document}